# Cupid is Doomed: An Analysis of the Stability of the Inner Uranian Satellites

Robert S. French[a] and Mark R. Showalter[b]

[a] SETI Institute, 189 Bernardo Ave., Mountain View, CA 94043, United States;
rfrench@seti.org
[b] SETI Institute, 189 Bernardo Ave., Mountain View, CA 94043, United States;
mshowalter@seti.org

## ABSTRACT

We have explored the stability of the inner Uranian satellites using simulations based on the most recent observational data. We find that, across a wide range of mass assumptions, the system is unstable, resulting in the eventual crossing of orbits and probable subsequent collision of moons. Cupid and Belinda are usually the first satellites to cross orbits, and they do so on a time scale of $10^3$–$10^7$ years. Cressida and Desdemona are generally the next pair to cross, on a time scale of $10^5$–$10^7$ years. We show that the crossing times are highly sensitive to initial conditions and that Cupid's instability is related to its resonant interactions with Belinda. We also show that a previously discovered power law, which relates orbit crossing time to satellite mass, is valid across a wide range of masses. We generalize the power law to handle two unstable orbital pairs with overlapping lifetimes and show that it can be used to extend the time span of studies of orbital stability in a computationally efficient manner. Our results suggest that the current Uranian satellite system is in transition and that the moons will continue to collide and reaccrete for the foreseeable future.

# 1. INTRODUCTION

Uranus possesses the most densely-packed system of satellites in the solar system. These 13 low-mass inner moons, with semi-major axes $a$ = 59,166–97,736 km (2.3–3.8 Uranian radii), are also very close to the planet. The first ten were discovered by Voyager 2 (Smith et al., 1986). Perdita was discovered later in archival Voyager 2 imagery (Karkoschka, 2001), and Cupid and Mab were discovered using the Hubble Space Telescope (HST) (Showalter and Lissauer, 2006). In addition, Uranus has five large "classical" satellites ($a$ = 129,848–583,520 km) as well as nine very distant "irregular" satellites ($a = 4.2 \times 10^6$–$2.1 \times 10^7$ km).

Duncan and Lissauer (1997, henceforth DL97) were the first to raise questions about the long-term stability of the inner satellite system. Using numerical integrations, they showed that the orbits of some moons could begin to cross on time scales as short as $\sim 10^6$ years. Such crossings are likely to lead to collisions because of unequal apsidal precession rates as well as the continued gravitational interactions between the bodies (Mikkola and Innanen, 1995). Multiple observations of the inner satellites provided by Voyager 2 (Smith et al., 1986), simultaneous analysis of data from HST and Voyager 2 (Owen and Synnott, 1987; Jacobson, 1998), and five years of more recent HST observations (Showalter and Lissauer, 2006; Showalter et al., 2008) have shown that the orbits of some of the inner satellites are variable over periods shorter than two decades; Showalter and Lissauer (2006) suggested that these variations may be a short-term manifestation of the predicted long-term instability.

DL97 studied the orbital dynamics of a subset of eight of the inner satellites along with the five classical satellites. They used the SWIFT simulator with the Regularized Mixed Variable Symplectic integrator (Levison and Duncan, 1994) and ran simulations until two orbits crossed or until a preset time limit was exceeded. Initial orbital state vectors for the inner satellites were derived from Voyager 2 images (Owen and Synnott, 1987), and those for the classical satellites were taken from unpublished measurements. Because the dynamical masses of the inner moons are unknown, DL97 used mass estimates by Lissauer (1995), which are based on estimated radii from unresolved images (Thomas et al., 1989) and the assumption that the densities are the same as that of Miranda, $\rho$ = 1.2 g/cm$^3$ (Jacobson et al., 1992).

Due to the compute time required to execute their simulations, DL97 were unable to explore a wide range of realistic mass estimates. Instead, they assumed a single set of masses and then introduced a *mass scaling factor*, $m_f$, which was uniformly applied to all satellites. Comparing the time to first orbit crossing, $t_c$, with $m_f$, they found that orbit crossing time generally decreased with increasing mass, presumably due to the stronger mutual perturbations. The results were well-described by a power law of the form $t_c = \beta m_f^{\alpha}$. Thus, by running a large number of simulations, DL97 were able to model $t_c(m_f)$ for $m_f > 1$ and then extrapolate to $t_c(1)$. They found that the five classical satellites, by themselves, were stable over a period ($\sim 2.5 \times 10^{17}$ years) much longer than the age of the solar system, while the inner satellites were stable over a much shorter period ($\sim 4$–100 million years), with either Cressida and Desdemona or Desdemona and Juliet crossing orbits within this time. When the oblateness of Uranus was taken into account, the classical satellites did not substantially affect the orbit-crossing times of the inner moons, presumably because the resulting orbital precession disrupted any secular resonances that would otherwise form.

Meyer and Lissauer (2005) used the Mercury hybrid symplectic integrator (Chambers, 1999) to simulate the same 13 satellites (eight inner and five classical) used by DL97. They explored a range of possible masses by assuming densities between 0.1 and 30 g/cm$^3$ and replicated the fundamental results of DL97. They showed that the first collision would generally occur in less than $3\times10^6$ years, with lower densities corresponding to longer times between collisions. By assuming that colliding bodies merge, they also continued the simulations past the first collision to explore system evolution further. The fact that satellites in the simulations of Meyer and Lissauer (2005) experienced collisions over time periods similar to the orbit crossings found by DL97 supports the argument that a collision will follow the crossing of two orbits.

Dawson et al. (2009, 2010) explored the short-term evolution of the orbits of the inner Uranian satellites by analyzing the effects of overlapping resonances from multiple pairs. They found that the evolution was very sensitive to the assumed masses and attributed this sensitivity to the dependence of the widths of the overlapping resonances on the masses. They also computed the Lyapunov characteristic exponent for each orbit and found that the Lyapunov time increased with decreasing mass, but they had concerns that the Lyapunov exponent was not a valid predictor of chaos in systems with strong resonances.

In this study, we take advantage of the improvements in computational speed since 1997 to expand upon the results of DL97. The paper is organized as follows. In Section 2, we discuss the simulation methodology, including the choice of simulator and computation environment and the selection of satellite masses and initial orbital parameters. In Section 3, we analyze the stability of the orbits under the various mass assumptions by direct integration. We examine the influence of the classical satellites on the inner system and explore the roles of individual inner moons. We also investigate the sensitivity of the simulations to small changes in initial conditions and explore the role of resonances in causing instability. In Section 4, we reproduce and expand upon the results of DL97 to verify the applicability of the power law over a larger range of mass factors and to extend the power law to handle the case of two independent, unstable systems with overlapping orbit crossing times. We then use the power law to predict the crossing times of the inner satellites using conservative density assumptions. In Section 5, we discuss the potential long-term evolution of the system. Finally, in Section 6 we discuss the implications of our results.

## 2. METHODOLOGY

*2.1. Simulator and Environment*

All simulations were performed using the SWIFT[1] simulator and the RMVS3 Regularized Mixed Variable Symplectic integrator (Levison and Duncan, 1994), which is based on the symplectic mapping method of Wisdom and Holman (1991). We applied minor modifications to the SWIFT driver to check for orbit crossings and to change the format of the output file, but the core algorithm is unchanged. The determination of the satellite masses and state vectors used to initialize the simulations are described in Section 2.2 and Section 2.3 respectively.

Simulations were run until any pair of satellites crossed orbits. We define the crossing time, $t_c$, as the time when the apoapsis of any one satellite becomes larger than the periapsis of the next satellite out from Uranus. For convenience, we will often refer to log $t_c$, the base 10 logarithm with $t_c$ measured in years. We used the traditional osculating element definitions for the eccentricity and semi-major axis to compute the periapses and apoapses. These are known to contain errors when the central body is oblate but are simple and fast to compute. Crossing times derived using the more accurate geometric orbital elements of Borderies-Rappaport and Longaretti (1994) were the same in 99% of all cases tested. Minor differences found in the remaining 1% of cases will not materially affect our conclusions.

---

[1] SWIFT is available at http://www.boulder.swri.edu/~hal/swift.html.

The majority of the runs, as well as the computationally-intensive procedure for setting up the initial conditions, were performed using the 1,160-processor supercomputer at the Centre for Astrophysics and Supercomputing at the Swinburne University of Technology. Additional simulations were performed on various desktop PCs running the 64-bit version of Windows 7. In all cases, SWIFT was compiled with the GNU 3.4 32-bit FORTRAN 77 compiler. A direct comparison of simulation results from the various systems showed no differences.

The numerical stability of the SWIFT simulator has already been well-established in previous studies (e.g., Levison and Duncan, 1994; DL97). However, to determine the appropriate simulation parameters, we characterized the sensitivity of orbit crossing time to integration step size. When the step size was kept sufficiently small, we found no systematic dependence on step size. After balancing the tradeoffs between precision and integration speed, we settled on a step size of 5% the period of the innermost moon, the same step size generally used by DL97. Time steps were 1447 seconds when Cordelia and Ophelia were included in the simulations, 1887 seconds otherwise.

*2.2. Satellites and mass estimates*

We explored numerous combinations of satellites, masses, and physical parameters in this study. Model "DL97(8J)" replicates the assumptions made by DL97 using the eight then-known satellites of the "Portia group" (Bianca, Cressida, Desdemona, Juliet, Portia, Rosalind, Belinda, and Puck). The assumed masses can be found in Table 1, and the initial state vectors can be found in DL97. For this model we also assumed the values from DL97 for the radius ($R_U$ = 26,200 km) and gravitational moments ($J_2$ = 3.34343×10$^{-3}$ and $J_4$ = –2.885×10$^{-5}$) of Uranus. Because DL97 did not publish their assumptions for the *GM* of Uranus, we assumed a value of 5793965.663939 km$^3$s$^{-2}$.[2,3]

Our remaining models were based on updated ephemerides and physical parameters. We also included Cordelia and Ophelia and the three inner moons discovered since 1997, Cupid, Perdita, and Mab. No dynamical mass estimates are available for these 13 inner bodies, and masses computed from volume and density estimates are subject to large uncertainties. For Cupid, Perdita, and Mab, we have used the radii found by Showalter and Lissauer (2006) and assumed a 20% uncertainty in albedo and a 20% uncertainty in disk-integrated photometry. This yields a 40% uncertainty in area and thus a ~20% uncertainty in radius. For the remaining 10 moons, we adopted the radius and uncertainty estimates of Karkoschka (2001a), which are based on a re-analysis of Voyager images in which the moons were resolved, albeit marginally in most

---

[2] From the ura083.bsp SPICE kernel, available at http://naif.jpl.nasa.gov/pub/naif/generic_kernels/spk/satellites/ura083.bsp.

[3] Throughout this paper, we present full 16-digit machine precision for all physical quantities that are used during simulation. This should not be construed to represent the level of precision actually available in the measurements, which is usually substantially less. Instead, because of the sensitivity of chaotic systems to small changes in input parameters, we provide the full precision so that our results may be reproduced.

**Table 1: Physical characteristics of the Uranian satellites from the DL97 models and models new to this work. The nominal radii and 1 σ uncertainties specified here are used to produce the mass estimates (*GM*) for the Inner(*sat+*) and Inner(*sat−*) models.**

| Satellite | DL97 $GM^a$ (km$^3$s$^{-2}$) | Radius (km) | $GM$ (km$^3$s$^{-2}$) | $GM$ $(r-1\,\sigma)^f$ (km$^3$s$^{-2}$) | $GM$ $(r+1\,\sigma)^f$ (km$^3$s$^{-2}$) |
|---|---|---|---|---|---|
| Cordelia |  | 21±3$^b$ | 0.002589112466095$^d$ | 0.001630461494684 | 0.003864797617028 |
| Ophelia |  | 23±4$^b$ | 0.003401547497568$^d$ | 0.001917581514410 | 0.005502807544558 |
| Bianca | 0.003569068560288 | 27±2$^b$ | 0.005502807544558$^d$ | 0.004368306044999 | 0.006818471432415 |
| Cressida | 0.012051400333440 | 41±2$^b$ | 0.019268353339353$^d$ | 0.016583906962132 | 0.022227898158064 |
| Desdemona | 0.008169458879880 | 35±4$^b$ | 0.011986631787478$^d$ | 0.008328717144741 | 0.016583906962132 |
| Juliet | 0.024856013187720 | 53±4$^b$ | 0.041621779139926$^d$ | 0.032891317624839 | 0.051774700889058 |
| Portia | 0.055853605391520 | 70±4$^b$ | 0.095893054299822$^d$ | 0.080375712941637 | 0.113289116721840 |
| Rosalind | 0.008169458879880 | 36±6$^b$ | 0.013043691957471$^d$ | 0.007548432845759 | 0.020712899728762 |
| Cupid |  | 9±2$^c$ | 0.000203807686836$^d$ | 0.000095893054300 | 0.000372109782137 |
| Belinda | 0.013210188827040 | 45±8$^b$ | 0.025475960854435$^d$ | 0.014161139590230 | 0.041621779139926 |
| Perdita |  | 13±3$^c$ | 0.000614218776375$^d$ | 0.000279571586880 | 0.001145125219860 |
| Puck | 0.152960081155200 | 81±2$^b$ | 0.148575803703066$^d$ | 0.137839695623702 | 0.159855398947324 |
| Mab |  | 12±3$^c$ | 0.000483099702129$^d$ | 0.000203807686836 | 0.000943554105720 |
| Miranda | 4.399977849623880 |  | 4.403988880239192$^e$ |  |  |
| Ariel | 90.29859336378001 |  | 86.48943821066345$^e$ |  |  |
| Umbriel | 78.20084149059601 |  | 81.48337213859010$^e$ |  |  |
| Titania | 235.2977975679480 |  | 228.6406014922988$^e$ |  |  |
| Oberon | 201.1019491218120 |  | 190.9467780172403$^e$ |  |  |

Sources: $^a$DL97; $^b$Karkoschka (2001); $^c$Showalter and Lissauer (2006) with uncertainties as described in Section 2.2; $^d$computed from the mean radii ($\rho = 1$ g/cm$^3$) as described in Section 2.2; $^e$SPICE kernel "ura083.bsp"; $^f$computed from the mean radii minus/plus 1 σ ($\rho = 1$ g/cm$^3$) as described in Section 2.2.

cases. These radii are generally larger than those assumed by DL97, which were based on the earlier work of Smith et al. (1986).

In all cases, we have assumed a "baseline" density of $\rho = 1$ g/cm$^3$ and computed the mass accordingly (Table 1). We call this set of moons and mass assumptions "Inner(baseline)". Despite using a lower density than the 1.2 g/cm$^3$ assumed by DL97, our baseline masses are on average ~55% larger.

To account for uncertainties in radius, we created 26 additional models using different sets of mass assumptions. For each of the 13 satellites, we defined mass sets "Inner(*sat−*)" and "Inner(*sat+*)", in which the radius of one satellite was either decreased or increased by 1 σ (Table 1). In addition, we created five models to account for the overall uncertainty in density. These are identified "Inner($\rho$ =*density*)", where *density* is given in units of g/cm$^3$. This yields 32 models in total.

For the models new to this paper, we assume a radius for Uranus of $R_U = 26{,}200$ km, gravitational moments for Uranus of $J_2 = 3.344247802666718 \times 10^{-3}$ and $J_4 = -2.772599495619087 \times 10^{-5}$, and a *GM* for Uranus of 5793965.663939 km$^3$s$^{-2}$ (all from the SPICE "ura083.bsp" kernel). The masses of the five classical satellites are known dynamically and are held fixed (Table 1).

*2.3. Initial orbital state vectors*

The orbits of the inner satellites are constrained by a series of unevenly spaced observations made over the past 24 years, first by Voyager 2 and more recently by HST. All relevant astrometric data has previously been collected and used to describe the motion of the Uranian moons within the SPICE information system,[4] which is maintained by the Navigation and

---

[4] For more information on SPICE, see http://naif.jpl.nasa.gov/naif/toolkit.html.

Ancillary Information Facility (NAIF) at NASA's Jet Propulsion Laboratory. SPICE kernel "ura083.bsp" contains a complete dynamical model for the motion of the classical satellites (see, for example, Jacobson et al., 1992; Jacobson, 1998). Kernel "ura091.bsp"[5] describes the motion of the inner satellites using simple precessing Keplerian ellipses (Jacobson, 1998).

For each model, we seek initial conditions for our SWIFT integrations in the form of a "state vector" (position and velocity) for each moon. Ideally, we would like to begin our integrations at the epoch of the Voyager 2 Uranus encounter in January 1986 and obtain results that agree with the kernels for the first 24 years over which astrometry has been available. Although the SPICE kernels provide these state vectors, we find that our SWIFT integrations diverge rapidly from the SPICE models. This is because SPICE kernel "ura091.bsp" does not incorporate the gravitational perturbations of each moon on the others; in effect, it treats each moon as a massless test particle.

In practice, because the moons interact measurably over the span of our analysis, we must derive a unique set of initial state vectors for each set of mass assumptions. Because our initial state vectors must be compatible with the observations over the 24-year period, we performed iterative orbital fits, seeking the initial state vectors that minimize the root-mean-square residuals between the positions of the moons as integrated by SWIFT and those as tabulated by the SPICE kernel. The fitting procedure consisted of a four-pass Powell optimization (Powell, 1964) over the six-dimensional vector of initial orbital elements for each satellite. This was sufficient to produce state vector components accurate to less than 1 mm or 1 mm/sec. The orbital elements were initialized for 00:00:00 UTC on January 1, 1986, just before the Voyager 2 encounter with Uranus, based on conversion from the state vectors provided by the SPICE library. For each step in the optimization, the current trial initial orbital elements were converted back to state vectors, which were used to initialize the SWIFT simulation. Although the optimization could have been performed using the state vectors directly, we found that optimizing the initial orbital elements resulted in much better convergence. We employed the geometric orbital elements of Borderies-Rappaport and Longaretti (1994), as implemented in closed form by Renner and Sicardy (2006), because they are not subject to the short-term oscillations seen in the osculating elements and caused by Uranus' oblateness.

Optimization was initially performed over 1.5 days, and the time period was progressively expanded to cover the entire 24-year period after seven passes. To optimize the state vectors for 13 satellites would require a Powell-style optimization over a 78-dimensional space (13 state vectors times six parameters per state vector). Such an optimization is computationally impractical. Thus, for each time span, each satellite was optimized separately, starting from the outermost and proceeding inwards, while the initial state vectors of the other satellites were held constant. During this process, the newly-determined initial state vectors of each satellite were used during the optimization of the satellites further inward. The outer-to-inner optimization direction was chosen because the more massive outer satellites tend to have greater influence on the generally less massive inner ones.

The resulting mean residuals were generally on the order of 10–10000 km RMS over the 24-year simulation period. Perdita often had an anomalously large residual of $\sim 10^5$ km because its 43:44 outer Lindblad resonance with Belinda is not described by the SPICE kernel. As expected, the magnitude of the residuals was sensitive to the exact masses chosen and varied from model to model for each moon. Of the unit-density models, the lowest mean residual (~1000 km) for all moons was found with the Inner(Mab−) model and the largest (~8800 km) was found with the Inner(Cordelia+) and Inner(Portia−) models. Due to the inherent instability of the system, we do not believe that the range of mean residuals reflects the physical accuracy of our various mass assumptions.

---

[5] Available at http://naif.jpl.nasa.gov/pub/naif/generic_kernels/spk/satellites/ura091.bsp.

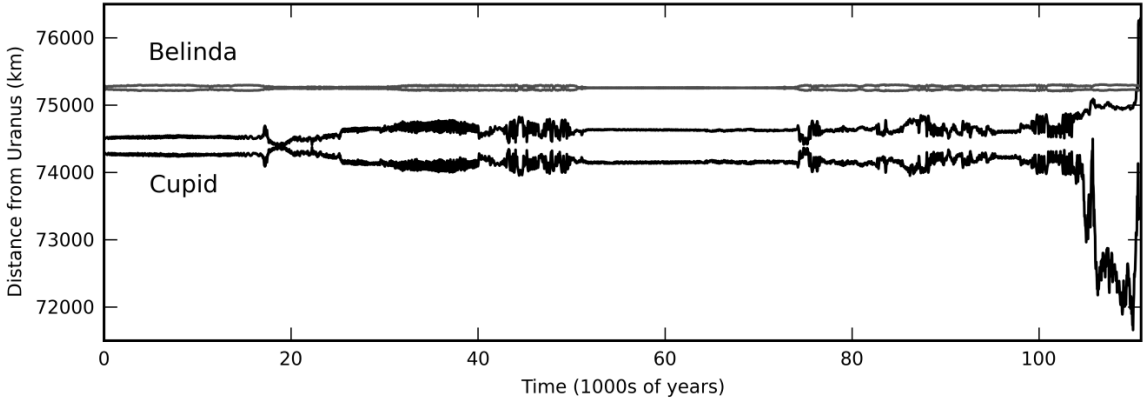

**Fig. 1: The apoapsis and periapsis of Cupid (bottom, black) and Belinda (top, gray) until orbit crossing for the Inner(baseline) model.**

## 3. STABILITY OF THE INNER SATELLITES

*3.1. Predictions of instability*

Gladman (1993) found that a system with a central body and two close planets on circular orbits will be Hill-stable (disallowing close approaches) as long as the initial fractional orbital separation of the two planets, $(a_2-a_1)/a_1$, is greater than $2.4(\mu_1 + \mu_2)^{1/3}$, where $\mu_1$ and $\mu_2$ are the mass ratios of the two planets to the central body. Chambers et al. (1996) expanded this result to multi-planet systems. They found that, across a wide range of assumptions, a system is always unstable when the initial orbital separations, $\Delta$, are less than 10 in units of mutual Hill radii, while systems with $\Delta \geq 10$ are decreasingly likely to be unstable.

In the Uranus system, the two closest pairs are Cupid–Belinda ($\Delta$=10.12) and Cressida–Desdemona ($\Delta$=11.78). Additional close pairs include Belinda–Perdita ($\Delta$=13.37), Juliet–Portia ($\Delta$=13.38) and Desdemona–Juliet ($\Delta$=18.39). For all other pairings, $\Delta$>30. Thus we expect the Cupid–Belinda–Perdita and Cressida–Desdemona–Juliet–Portia systems to be particularly unstable, and these are the primary focus of our investigation.

*3.2. Simulation results*

From direct integrations, we have determined the orbit crossing times for our 32 models (Table 2). We find that Cupid and Belinda are almost always the first to cross. For the 27 models that assume $\rho = 1$ g/cm$^3$, log $t_c$ = 4.1 to 5.7. Allowing a wider range of densities ($\rho$ = 0.5–3.0 g/cm$^3$) expands the range of log $t_c$ to 3.1 to 6.2, in all cases resulting in an orbit crossing in less than ~$10^6$ years. As an example of orbital evolution, the apoapse and periapse of Cupid and Belinda from a simulation of the Inner(baseline) model are shown until orbit crossing in Fig. 1. During most of the simulation, the eccentricity of both Cupid and Belinda vary in an irregular manner. However, near the end, the semi-major axis of Cupid begins to vary dramatically, along with increased eccentricity, until orbit crossing occurs. These variations are suggestive of chaos.

To explore the accuracy of our crossing times and the nature of the instability, it is worthwhile to analyze the sensitivity of the results to small variations in initial conditions. To do this, we started with the Inner(baseline) model as a reference. We then varied the initial conditions by moving, alone or in combination, the spatial X, Y, and Z components of the initial state vector by –2, –1, 0, +1, and +2 mm, resulting in 124 additional models. The result was a dramatic spread in crossing times, with a minimum of log $t_c$ = 4.2 and a maximum of log $t_c$ = 5.9, a factor of ~50. Values of log $t_c$ are approximately normally distributed around a mean of 5.0 with a standard deviation of 0.4 (a factor of 2.5). We found no correlation between the crossing times and the change made in the X, Y, or Z direction. In addition to the spread in crossing times, some simulations had different pairs of satellites (Rosalind and Cupid, or Belinda and Perdita) crossing first as well.

Table 2: Models new to this work with crossing times from single simulations and first satellites to cross. $t_c$ is in years.

| Model | log $t_c$ | Crossing Satellites |
| --- | --- | --- |
| Inner($\rho$ = 0.5) | 6.2 | Cupid–Belinda |
| Inner($\rho$ = 0.7) | 5.3 | Cupid–Belinda |
| Inner(baseline) | 5.0 | Cupid–Belinda |
| Inner($\rho$ = 1.5) | 3.4 | Cupid–Belinda |
| Inner($\rho$ = 2.0) | 3.1 | Cupid–Belinda |
| Inner($\rho$ = 3.0) | 3.4 | Cressida–Desdemona |
| Inner(Cordelia–) | 4.7 | Cupid–Belinda |
| Inner(Cordelia+) | 4.9 | Cupid–Belinda |
| Inner(Ophelia–) | 5.1 | Cupid–Belinda |
| Inner(Ophelia+) | 5.4 | Cupid–Belinda |
| Inner(Bianca–) | 5.4 | Cupid–Belinda |
| Inner(Bianca+) | 4.8 | Cupid–Belinda |
| Inner(Cressida–) | 5.0 | Cupid–Belinda |
| Inner(Cressida+) | 5.5 | Cupid–Belinda |
| Inner(Desdemona–) | 5.5 | Cupid–Belinda |
| Inner(Desdemona+) | 4.8 | Cupid–Belinda |
| Inner(Juliet–) | 5.0 | Cupid–Belinda |
| Inner(Juliet+) | 5.6 | Cressida–Desdemona |
| Inner(Portia–) | 4.6 | Cupid–Belinda |
| Inner(Portia+) | 5.1 | Cupid–Belinda |
| Inner(Rosalind–) | 4.8 | Cupid–Belinda |
| Inner(Rosalind+) | 4.8 | Cupid–Belinda |
| Inner(Cupid–) | 5.0 | Cupid–Belinda |
| Inner(Cupid+) | 5.3 | Cupid–Belinda |
| Inner(Belinda–) | 5.7 | Cupid–Belinda |
| Inner(Belinda+) | 4.1 | Cupid–Belinda |
| Inner(Perdita–) | 5.4 | Cupid–Belinda |
| Inner(Perdita+) | 4.4 | Cupid–Belinda |
| Inner(Puck–) | 4.6 | Cupid–Belinda |
| Inner(Puck+) | 4.6 | Cupid–Belinda |
| Inner(Mab–) | 4.7 | Cupid–Belinda |
| Inner(Mab+) | 5.6 | Cupid–Belinda |

We also started with the Inner(baseline) model and varied the mass of Cupid by –6.2×10$^5$ kg to +6.2×10$^5$ kg in increments of 10$^4$ kg (a change of approximately one part in 10$^{11}$). We vary the mass of Cupid because it is the least massive satellite, and if small changes in Cupid's mass cause noticeable changes in the stability of the entire Uranian system, it is a strong statement about the system's sensitivity to initial conditions. Again the resulting crossing times showed a wide spread (a factor of ~130) with no apparent correlation to mass, as well as a variety of crossing satellites, with the same mean and standard deviation as the position variation.

There is little doubt that the simulations are extremely sensitive to initial conditions. This strong dependence on very tiny changes in initial conditions is one of the common characteristics of chaos. Thus, all individual crossing times discussed in this paper must be considered samples of a statistical distribution. For each of our key conclusions below, we have run 10–20 simulations using mm-scale variations in the starting conditions in order to characterize the time scales and their uncertainties more precisely. As the fractional uncertainty of the standard deviation of $N$ measurements is $(2N–2)^{-1/2}$, even this small number of simulations provides standard deviations with an accuracy of 16–24%.

DL97 found that the addition of the five classical satellites had little effect on their results as long as Uranus was assumed to be oblate (an assumption we make throughout this paper). To test this hypothesis in our new models, we created the "Combined" model, which uses the mass assumptions of Inner(baseline) for the inner satellites and the dynamically-measured masses of the five classical satellites; the initial orbital state vectors are recomputed for the complete set of 18 moons. The resulting log crossing time of 5.1±0.4 (with Cupid and Belinda continuing to cross first in most cases) is consistent with the Inner(baseline) model's 5.0±0.4. Thus it appears that the classical satellites do not fundamentally change the instability of the inner satellites as measured by time to first orbit crossing, and we will ignore the classical satellites for the rest of this paper.

**Table 3:** Orbit crossing time of the Inner(baseline) model with individual satellites removed. $t_c$ is in years. Crossing times without uncertainties are the result of a single simulation; crossing times with uncertainties are the mean and standard deviation of 10–20 simulations. In the cases with multiple simulations the crossing satellites are the pair that most commonly cross first.

| Satellite removed | log $t_c$ | Crossing Satellites |
|---|---|---|
| None | 5.0±0.4 | Cupid–Belinda |
| Cordelia | 4.9 | Cupid–Belinda |
| Ophelia | 5.1 | Cupid–Belinda |
| Bianca | 4.4 | Cupid–Belinda |
| Cressida | 4.8 | Rosalind–Cupid |
| Desdemona | 4.8 | Cupid–Belinda |
| Juliet | 4.6 | Cupid–Belinda |
| Portia | 4.7 | Cupid–Belinda |
| Rosalind | 5.0 | Cupid–Belinda |
| Cupid | 5.6±0.4 | Belinda–Perdita |
| Belinda | 6.2 | Cressida–Desdemona |
| Perdita | 6.2±0.2 | Cressida–Desdemona |
| Puck | 5.4 | Rosalind–Cupid |
| Mab | 4.6 | Cupid–Belinda |

As Cupid, which was not yet discovered at the time of DL97, apparently has a particularly unstable orbit, we conducted a simulation with Cupid removed. The orbital fits were not recomputed; we used the initial state vectors from the Inner(baseline) model, which is reasonable because of Cupid's low mass. In these simulations, Belinda and Perdita usually cross first, with log $t_c$ = 5.6±0.4, approximately four times longer than when Cupid is included (Table 3).

For completeness, we further explored the effect of removing each of the other inner satellites on the crossing time (Table 3) using single simulations, again without recomputing the initial state vectors. In general, the removal of a single satellite does not dramatically change the time of first orbit crossing, and Cupid and Belinda remain the first pair of satellites to cross. That the removal of Belinda, a relatively massive moon that interacts strongly with Cupid, results in Cupid's orbit becoming more stable and Cressida and Desdemona crossing first is not surprising (Section 3.3). However, when Perdita, a very low mass satellite not modeled by DL97, is removed, Cressida and Desdemona are again the first satellites to cross, and an ensemble of simulations yields a log crossing time of 6.2±0.2. This suggests that the presence of Perdita destabilizes the Cupid–Belinda system. The effect of Perdita will be discussed further in Section 4.4.

Although DL97 were not able to find the crossing time for many of their models with direct integration, by using the power law their predicted log crossing time for Cressida and Desdemona was 6.6±0.1 for the 8J model. The shorter crossing times in our simulations with Belinda or Perdita removed can be accounted for by our ~55% larger baseline masses. Because our simulations stop at the first orbit crossing, the instability of Cupid caused by Belinda and Perdita effectively "hides" that of Cressida and Desdemona. This will be explored further in Section 4.2 and Section 5.

**Table 4: Correlated semi-major axes in a simulation of the Inner(baseline) model and associated resonances.**

| Inner Satellite | Distance from ILR | Outer Satellite | Correlation Mean | σ | Adjacent Resonance separation (km) |
|---|---|---|---|---|---|
| Bianca | 4.2 km from 16:15 | Cressida | –0.14 | 0.24 | 154 |
| Cressida | 0.9 km from 47:46 | Desdemona | –0.93 | 0.05 | 19 |
| Juliet | 1.9 km from 51:49 | Portia | –0.95 | 0.05 | 34 |
| Belinda | 0.2 km from 44:43 | Perdita | –0.63 | 0.33 | 26 |
| Belinda | 10.1 km from 11:9 | Puck | –0.26 | 0.14 | 925 |
| Cupid | 3.9 km from 58:57 | Belinda | –0.36 | 0.26 | 15 |

We predicted in Section 3.1 that the subsets of moons {Cupid, Belinda, Perdita} and {Cressida, Desdemona, Juliet, Portia} would be unstable based on the criterion of Gladman (1993), a result that is supported by the simulations discussed above. Examining the instability of these two groups of satellites in isolation permits us to determine what effect, if any, the remaining satellites have on their stability. It is worth noting that Juliet and Portia, despite their predicted instability, almost never cross in our simulations. The reason for this is unclear, although it is probably related to their near-resonant interaction discussed in Section 3.3. As a result of this observation we removed Portia from our isolation studies.

A simulation containing only Cupid, Belinda, and Perdita, using the baseline mass assumptions with a new orbital fit, resulted in a log crossing time of 6.2±0.4 with Cupid and Belinda usually crossing first. While this is approximately an order of magnitude longer than any of our 13-satellite models, most of which also have Cupid and Belinda crossing first, it is nevertheless clear that the Cupid–Belinda–Perdita system is unstable on astronomically short timescales even in isolation.

Likewise, a simulation containing only Cressida, Desdemona, and Juliet (also using the baseline mass assumptions with a new orbital fit) resulted in a log crossing time of 6.9±0.2 with Cressida and Desdemona usually crossing first. This is noticeably longer than the crossing time of any of our 13-satellite models, and in particular is longer than the log($t_c$) = 6.2±0.2 from the Inner(baseline) model with Perdita deleted, which is our most similar model that results in a crossing of Cressida and Desdemona. Like the Cupid–Belinda–Perdita system, the Cressida–Desdemona–Juliet system is apparently unstable on its own, although also on a much longer timescale. In both cases, the addition of other satellites decreases the overall stability and thus decreases the crossing time.

*3.3. Resonant interactions*

Much of the orbital instability of the inner Uranian satellites can be attributed to resonant interactions. Belinda and Perdita are the best-known example of an interacting pair in this system, with Perdita located at Belinda's 43:44 outer Lindblad resonance (OLR), leading to orbital libration (Showalter and Lissauer, 2006).

Other near-resonances also play a role. We identified these interactions by solving for the correlation coefficient of semi-major axes between every pair of satellites in the system. We performed these calculations within a sliding window so that we could detect changes over time, indicated by a non-zero standard deviation of the mean correlation coefficients. In each case where we found a statistically significant correlation, the pair of satellites was also related by a near-resonance (Table 4).

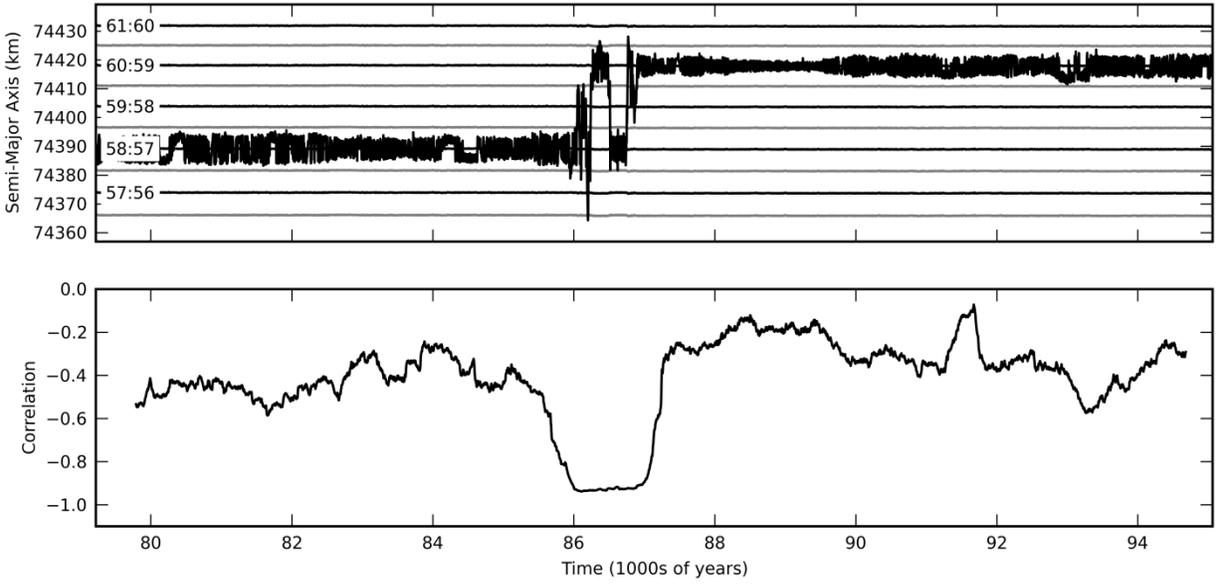

**Fig. 2: Resonant interactions of Cupid and Belinda. Top: The semi-major axis of Cupid overlaid on the first- (black) and second-order (gray) inner Lindblad resonances of Belinda for the Inner(baseline) model. Bottom: The correlation of the semi-major axis of Cupid with the semi-major axis of Belinda using a sliding window.**

Two of these pairs, Juliet–Portia and Cupid–Belinda, are particularly interesting. Juliet and Portia have the strongest consistent anti-correlation. This is unexpected because, in our simulations, Juliet and Portia almost never cross orbits. Visual inspection of the orbital elements shows that whenever Juliet approaches the 51:49 Inner Lindblad Resonance (ILR) of Portia, it is repelled. This prevents Juliet and Portia from getting close enough to cross orbits.

Other than Belinda and Perdita, Cupid and Belinda have the most varying interaction. The uncertainty in the semi-major axis of Cupid's orbit is only ~100 m, but the uncertainty in the eccentricity is nearly 60% (Showalter and Lissauer, 2006), making it difficult to place Cupid precisely in a resonance of Belinda. Nevertheless, it is enlightening to observe the orbital interaction of these satellites over time, both before and after the onset of significant instability that starts around 85,000 years (Fig. 1). Fig. 2 shows the semi-major axis of Cupid plotted against the first- and second-order ILRs of Belinda during this time. Cupid tends to maintain a stable orbit at one of Belinda's ILRs, but it is occasionally kicked out of a resonance and settles into a different resonance. During the period when Cupid is moving between resonances, the semi-major axes of Cupid and Belinda become highly anti-correlated as they exchange angular momentum. However, due to the much larger mass of Belinda, its orbit is mostly unaffected while Cupid undergoes major orbital changes.

Referring back to Fig. 1, the extreme instability that begins around 103,000 years and continues until orbit crossing is also characterized by strong anti-correlation. The behavior of Cupid during a portion of this time period is shown in Fig. 3. Cupid rapidly hops between many first- and second-order resonances, interrupted by occasional, brief periods of stability. It is clear that this resonant interaction is the primary cause of Cupid's instability and the eventual orbital crossing.

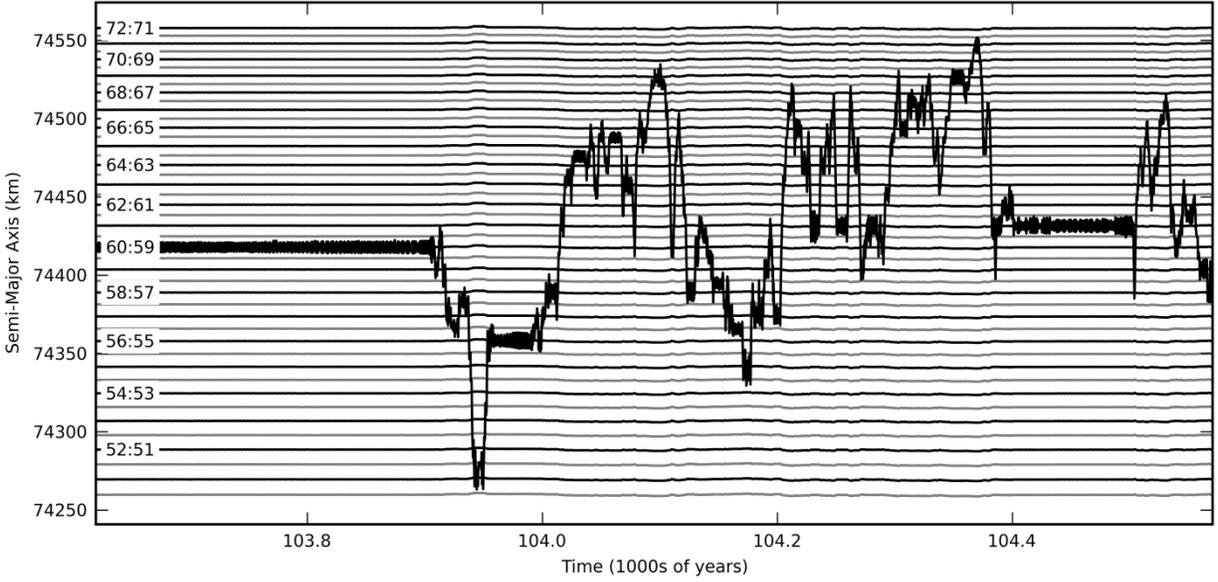

**Fig. 3: Resonant interactions of Cupid and Belinda (detail). The semi-major axis of Cupid overlaid on the first- (black) and second-order (gray) inner Lindblad resonances of Belinda for the Inner(baseline) model. The simulation is shown starting at the time when Cupid becomes completely anti-correlated with Belinda just before the orbits cross.**

## 4. THE POWER LAW

*4.1. The validity of the power law*

DL97 found that the time of orbit crossing, $t_c$, could be predicted by running many simulations with the satellite masses increased by mass factors, $m_f$, and fitting the results to a power law of the form $t_c(m_f) = \beta m_f^{\alpha}$. The time of predicted orbit crossing is then easy to find, because when $m_f = 1$, $t_c = \beta$. As simulation time decreases as approximately the fourth power of the mass factor for the Uranus system, this technique allows the study of models with low mass assumptions that would otherwise be computationally impractical. Due to computational limitations, DL97 relied heavily on the power law to make their prediction that the Uranian satellite system was unstable in the time span of 4–100 million years. However, they were able to verify the operation of the power law with a simulation at $m_f = 1$ for only a small subset of their models. It is thus reasonable for us to ask whether the power law is truly applicable for their most complete models and whether it is applicable to our new models as well.[6]

The simulations of the DL97(8J) model performed by DL97 covered a mass factor range of ~1.3–40. We started our exploration by reproducing their results. We ran a series of simulations with varying mass factors $2^{n/10}$ for integer $n = -1$ to 50. In keeping with the technique used by DL97, the initial orbital state vectors were not recomputed based on the new masses despite the greater perturbations. Recomputing the state vectors, which is a very time-consuming process, would reduce or eliminate the benefit of using the power law, which is to save simulation time by using multiple, shorter simulations to predict the results that would be found during longer simulations. For each simulation, we recorded the earliest time, $t_c(m_f)$, that any pair of satellites crossed orbits. Our power law fit parameters, $\alpha = -4.1$ and $\log \beta = 6.8$, are consistent with those found by DL97 using a slightly narrower range of $m_f$, $\alpha = -4.1 \pm 0.1$ and $\log \beta = 6.6 \pm 0.1$, and also consistent with our result for $m_f = 1$ by direct integration, $\log t_c = 7.0$. Although the power law seems to work for chaotic systems in general, it may be possible that it breaks down for sufficiently small masses. In particular, Smith and Lissauer (2009) found that, in an investigation of Earth-mass planets orbiting a Sun-like star, a similar power law became flat for mutual Hill radii separations less than ~3. However, we have not seen evidence for such a flattening and we do not extrapolate our power law to separations that small. Thus, given the lack of compelling evidence to the contrary, we will proceed under the assumption that the power law remains valid for arbitrarily small masses for the Uranian system.

---

[6] Note the important but perhaps subtle distinction between the mass factor and the density. When we adjust $m_f$, we leave the initial state vectors unchanged. Thus, the Inner(baseline) model with $m_f = 2$ differs from the Inner($\rho$=2) model in the values of the initial conditions, even though the assumed masses of the moons are the same. This is consistent with the prior work of DL97 and others when they employed the power law.

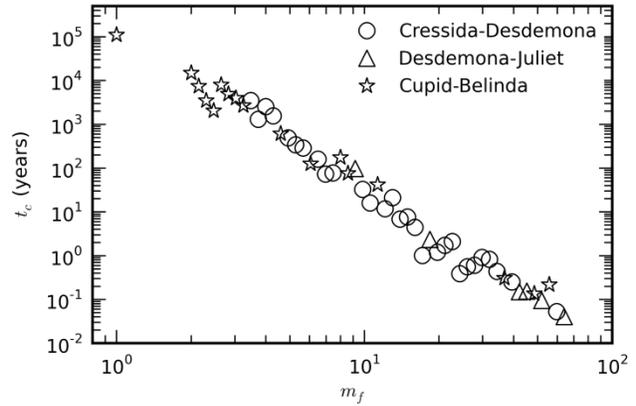

**Fig. 4:** Time of first crossing and first pair of satellites to cross at various mass factors for the Inner(baseline) model.

*4.2. Overlapping power laws*

In both our models and the models of DL97, the particular pair of satellites to cross first is dependent upon the mass factor. For the Inner(*sat*±) models, Cupid and Belinda are usually the first pair to cross at low mass factors. However, at higher mass factors, Cressida, Desdemona, and Juliet become relatively more unstable and one of these adjacent pairs usually crosses first. This can be seen in Fig. 4 for the Inner(baseline) model, although a similar trend is observed for the other models as well.

In their implementation of the power law, DL97 measured the time until the first pair of satellite orbits intersected, no matter which pair. Most often, it was Cressida and Desdemona, although occasionally it was Desdemona and Juliet. The challenge we have in applying the power law to our models is that we include Cupid, which we find to have a substantially shorter lifetime under many circumstances. To account for this, we generalize the power law concept to one in which each pair of adjacent satellites has a crossing time that can be modeled by a different power law. For simplicity, we assume that the distributions are Gaussians in $\log t_c$ (as implied by Section 3.2) with standard deviation $\sigma$, where each mean is defined by a power law with a different slope $\alpha$ and intercept $\beta$. This view of the crossing events as a set of independent random variables (RVs) with overlapping probability density functions also provides a natural framework within which we can understand a number of our anomalous integrations, such as the relatively rare cases where the first crossing involves Desdemona and Juliet or Rosalind and Cupid. Although the concept is valid for an arbitrary number of overlapping power laws, for practical reasons we will limit our analysis to two, representing satellite pairs #1 and #2.

In a given integration, the first crossing time can be regarded as the minimum of two independent RVs. Fig. 5 shows a Monte Carlo simulation of how the mean value of the minimum relates to the mean and standard deviation of the two distributions. When the mean of one distribution is much smaller than that of the other, our measurements are consistent with the smaller mean by itself. However, if the means of the two RVs are comparable, then our crossing time becomes a distinctly biased measure of the expected crossing time for either RV alone; the measurements are low by about 0.5 σ in the case where the means and standard deviations are equal. The figure shows a distinct bend in observed value in this region, but in practice the scatter among our simulations can make the bend much less obvious. An alternative (and intuitive) way to identify the rough location of the bend is the place where the most common first crossing transitions from one pair to the other.

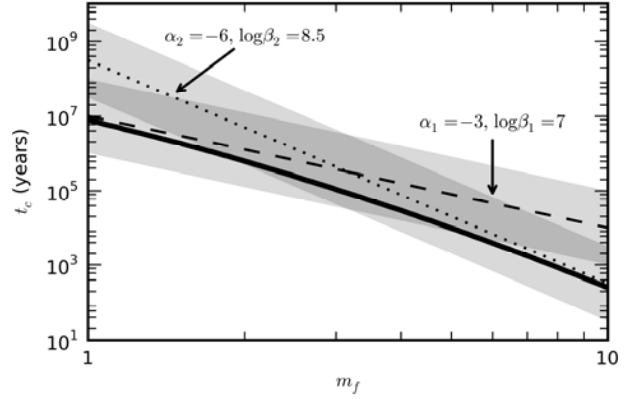

**Fig. 5: Monte Carlo simulation showing the measured minimum time of crossing (solid) vs. two random variables. Each random variable has a mean defined by a power law (dashed and dotted) and a standard deviation σ=1.0 (shaded). The maximum difference between the measured minimum and the minimum power law mean is ~0.55 (a factor of ~3.5).**

With a sufficient number of simulations at different mass factors, we can roughly determine the power laws of the satellite pairs. Our process for determining the two pairs of power law parameters ($\alpha_1$, $\beta_1$, $\alpha_2$, and $\beta_2$) along with the standard deviations ($\sigma_1$ and $\sigma_2$) of the associated RVs begins by running a set of simulations with $m_f$ = 2–64 in steps of $2^{n/10}$ for integer $n$. We also include a single simulation with $m_f = 1$, although the lengthy simulation times required for such small mass factors preclude us from running additional simulations with $m_f < 2$. For simplicity, we discard the results from those rare simulations in which neither pair #1 nor pair #2 crosses first. We next make a visual determination of the approximate mass factor where the crossing dominance switches from pair #1 to pair #2. Given this value, which we will call $m_{f,intersect}$, we determine a series of four metrics from our ensemble of simulations. These are, for the simulations with $m_f < m_{f,intersect}$: the fraction of simulations where pair #1 cross first, the slope $\alpha_1$ and intercept $\beta_1$ of the best fit line to the pair #1 crossings, and the standard deviation $\sigma_1$ of the pair #1 crossings about this line:

$$\alpha_1 = \frac{N \sum_{i=1}^{N} lm_{f,i} lt_{c,i} - \sum_{i=1}^{N} lm_{f,i} \sum_{i=1}^{N} lt_{c,i}}{\xi} \quad (1)$$

$$\log \beta_1 = \frac{\sum_{i=1}^{N} lm_{f,i}^2 \sum_{i=1}^{N} lt_{c,i} - \sum_{i=1}^{N} lm_{f,i} \sum_{i=1}^{N} lm_{f,i} lt_{c,i}}{\xi} \quad (2)$$

$$\sigma_{1,measured} = \sqrt{\frac{1}{N} \sum_{i=1}^{N} \left( \log \beta_1 + \alpha_1 lm_{f,i} - lt_{c,i} \right)^2} \quad (3)$$

where $lm_{f,i} \equiv \log m_{f,i}$ and $lt_{c,i} \equiv \log t_{c,i}$, in each case limited to those simulations where pair #1 cross first, and

$$\xi = N\sum_{i=1}^{N} lm_{f,i}^2 - \left(\sum_{i=1}^{N} lm_{f,i}\right)^2. \tag{4}$$

We also compute the uncertainties in $\alpha_1$ and $\beta_1$, $\sigma_{\alpha,1}$ and $\sigma_{\log \beta,1}$:

$$\sigma_{\alpha,1} = \sqrt{\frac{N}{N-2} \frac{\sum_{i=1}^{N}\left(\log \beta_1 + \alpha_1 lm_{f,i} - lt_{c,i}\right)^2}{\xi}} \tag{5}$$

$$\sigma_{\log \beta,1} = \sqrt{\frac{1}{N-2} \frac{\sum_{i=1}^{N} lm_{f,i}^2}{\xi}}. \tag{6}$$

For the simulations with mass factors greater than $m_{f,intersect}$, the metrics (and their uncertainties) are repeated for pair #2 crossings, yielding eight metrics and four uncertainties in total. Although many different metrics could be used, we find these to be sufficient to produce robust results.

Given these means and standard deviations of the metrics, we then seek the power law parameters and their uncertainties that, when run through MC simulations, produce the same distribution of values for the metrics. To determine the mean and standard deviation of the power law parameters, we perform the entire optimization procedure multiple times while perturbing the target metrics. For each optimization, the slope of the pair #1 crossings is chosen from a normal distribution with mean $\alpha_1$ and standard deviation $\sigma_{\alpha,1}$. Likewise, the intercept is chosen from a normal distribution with mean $\log \beta_1$ and standard deviation $\sigma_{\log \beta,1}$. A similar procedure is used for the pair #2 slope and intercept.

For each set of perturbed metrics, we use a Nelder-Mead simplex algorithm (Nelder and Mead, 1965) to find the optimal power law parameters that minimize the residuals found by computing the sum-of-squares of the differences between these eight metrics and the eight metrics from the Monte Carlo simulation. We weight the metrics equally. To avoid problems with local minima, we run the optimization multiple times using different initial values of the power law parameters and then choose the result with the lowest final residual.

We now apply our procedure to the specific case of Cupid–Belinda (taking the place of pair #1) and Cressida–Desdemona (pair #2). The values chosen for $m_{f,intersect}$ for the 27 unit-density models are shown in Table 5 (the exact value chosen does not have a strong effect on the end result) along with the results of the optimization procedure. Plots of selected solutions are shown in Fig. 6. In all cases, $\sigma_{cupid}$ and $\sigma_{cressida}$ are approximately 0.3 and are not included in the table. Note that these uncertainties are similar to those derived from perturbing the initial state vectors (Section 3.2), increasing confidence in our methodology.

Comparing Table 2 and Table 5, we find little difference between the crossing times found by direct integration and $\beta_{cupid}$. This shows that Cupid–Belinda is sufficiently unstable at our mass assumptions that the presence of Cressida–Desdemona does not bias the crossing times. We can also compare $\beta_{cressida}$ with the crossing time found with Perdita removed (Table 3) and find that our power law predictions (log $\beta_{cressida}$ = 5.4±0.2) are consistent with the results of direct integration (6.2±0.2). Thus we feel confident in our methodology and can apply it to cases where the assumed masses are small enough to make direct integration impractical.

**Table 5: Overlapping power law Monte Carlo fits for Cupid–Belinda and Cressida–Desdemona with extrapolation to $\rho = 0.5$ g/cm$^3$. $\sigma_{cupid}$ and $\sigma_{cressida}$ are ~0.3 in all cases.**

| Model | $m_{f,intersect}$ | $\alpha_{cupid}$ | log $\beta_{cupid}$ | $\alpha_{cressida}$ | log $\beta_{cressida}$ | log $t_{c,cupid}$ ($\rho = 0.5$) | log $t_{c,cressida}$ ($\rho = 0.5$) |
|---|---|---|---|---|---|---|---|
| Inner(baseline)   | 5.0  | −3.1±0.3 | 5.0±0.2 | −3.7±0.1 | 5.4±0.2 | 5.9±0.2 | 6.5±0.2 |
| Inner(Cordelia−)  | 4.0  | −2.5±0.3 | 4.7±0.1 | −3.7±0.2 | 5.6±0.2 | 5.4±0.2 | 6.7±0.3 |
| Inner(Cordelia+)  | 7.0  | −2.8±0.3 | 4.9±0.1 | −4.1±0.2 | 5.9±0.2 | 5.7±0.2 | 7.2±0.2 |
| Inner(Ophelia−)   | 8.0  | −3.2±0.2 | 5.0±0.2 | −3.3±0.2 | 5.1±0.2 | 6.0±0.2 | 6.1±0.2 |
| Inner(Ophelia+)   | 4.0  | −3.4±0.5 | 5.2±0.2 | −4.0±0.2 | 5.7±0.2 | 6.2±0.3 | 6.9±0.2 |
| Inner(Bianca−)    | 8.0  | −3.4±0.2 | 5.0±0.1 | −3.7±0.2 | 5.3±0.2 | 6.1±0.2 | 6.4±0.2 |
| Inner(Bianca+)    | 6.0  | −2.6±0.3 | 4.6±0.2 | −4.0±0.3 | 5.9±0.4 | 5.4±0.2 | 7.1±0.4 |
| Inner(Cressida−)  | 5.0  | −3.1±0.3 | 5.0±0.2 | −4.2±0.2 | 6.1±0.2 | 6.0±0.3 | 7.3±0.2 |
| Inner(Cressida+)  | 3.0  | −3.5±1.1 | 5.5±0.4 | −4.2±0.4 | 5.9±0.4 | 6.5±0.7 | 7.2±0.4 |
| Inner(Desdemona−) | 5.0  | −3.3±0.5 | 5.4±0.2 | −4.0±0.2 | 5.8±0.2 | 6.4±0.3 | 7.0±0.2 |
| Inner(Desdemona+) | 5.0  | −2.8±0.5 | 4.9±0.3 | −3.8±0.3 | 5.5±0.3 | 5.7±0.4 | 6.7±0.3 |
| Inner(Juliet−)    | 9.0  | −3.0±0.5 | 5.1±0.2 | −3.8±0.4 | 5.9±0.4 | 6.0±0.3 | 7.1±0.4 |
| Inner(Juliet+)    | 4.0  | −3.6±0.6 | 5.4±0.3 | −4.0±0.3 | 5.6±0.3 | 6.5±0.4 | 6.8±0.3 |
| Inner(Portia−)    | 6.0  | −2.5±0.4 | 4.6±0.3 | −3.8±0.2 | 5.7±0.2 | 5.3±0.3 | 6.8±0.2 |
| Inner(Portia+)    | 3.5  | −2.9±0.9 | 4.8±0.3 | −3.9±0.2 | 5.4±0.2 | 5.7±0.5 | 6.5±0.2 |
| Inner(Rosalind−)  | 4.0  | −2.4±0.5 | 4.8±0.2 | −3.7±0.2 | 5.4±0.2 | 5.5±0.3 | 6.5±0.2 |
| Inner(Rosalind+)  | 6.0  | −3.0±0.3 | 4.9±0.2 | −3.9±0.3 | 5.6±0.3 | 5.8±0.2 | 6.8±0.3 |
| Inner(Cupid−)     | 4.0  | −2.8±0.5 | 4.8±0.2 | −3.9±0.1 | 5.6±0.2 | 5.6±0.3 | 6.8±0.2 |
| Inner(Cupid+)     | 3.0  | −3.7±0.3 | 5.4±0.1 | −4.2±0.2 | 5.8±0.2 | 6.5±0.2 | 7.0±0.2 |
| Inner(Belinda−)   | 3.5  | −4.2±0.6 | 5.9±0.3 | −4.5±0.3 | 6.1±0.3 | 7.2±0.4 | 7.5±0.3 |
| Inner(Belinda+)   | 10.0 | −2.9±0.3 | 4.6±0.2 | −3.6±0.5 | 5.6±0.7 | 5.5±0.2 | 6.7±0.8 |
| Inner(Perdita−)   | 3.0  | −3.5±0.6 | 5.4±0.3 | −4.1±0.2 | 5.9±0.4 | 6.4±0.4 | 7.1±0.4 |
| Inner(Perdita+)   | 4.0  | −2.0±0.4 | 4.4±0.2 | −3.9±0.2 | 5.6±0.2 | 5.0±0.3 | 6.8±0.2 |
| Inner(Puck−)      | 8.0  | −2.6±0.3 | 4.8±0.2 | −4.2±0.3 | 6.0±0.3 | 5.6±0.2 | 7.3±0.3 |
| Inner(Puck+)      | 3.5  | −2.8±0.6 | 4.7±0.2 | −3.8±0.2 | 5.4±0.2 | 5.6±0.4 | 6.5±0.2 |
| Inner(Mab−)       | 4.0  | −2.9±0.4 | 4.8±0.2 | −3.9±0.1 | 5.6±0.2 | 5.7±0.3 | 6.8±0.2 |
| Inner(Mab+)       | 3.5  | −3.6±0.4 | 5.3±0.2 | −3.9±0.2 | 5.6±0.3 | 6.4±0.2 | 6.7±0.2 |

*4.3. Densities and Lifetimes*

The densities of the inner Uranian satellites are unknown, but it is possible to set reasonable limits. Miranda, the innermost of the classical Uranian satellites, is relatively large (235 km radius) and has a density of 1.2 g/cm$^3$ (Jacobson et al., 1992). This value, used by DL97 for all the inner satellites, is a conservative upper limit. However, the inner satellites are much smaller, more irregularly shaped, and are likely to be loose rubble piles. In this way they are more similar to the small icy satellites of Jupiter and Saturn than they are to Miranda. Amalthea, the largest of Jupiter's inner moons, has a radius (83.5 km) similar to that of Portia and Puck and a density of 0.857 g/cm$^3$ (Anderson et al., 2005). Saturn's innermost satellites Pan, Atlas, Prometheus, Pandora, and Epimetheus, with sizes comparable to the smaller of the inner Uranian satellites, have densities ranging from 0.42 to 0.64 g/cm$^3$, with a general trend of increasing density with increasing size (Thomas, 2010). Finally, Neptune's satellites Galatea and Despina may have densities of 0.4 to 0.8 g/cm$^3$ (Zhang and Hamilton, 2008). It is worth noting that the low albedo of the Uranian moons suggests that they are rocky and probably denser than the icy moons discussed here. As such, we choose 0.5 g/cm$^3$ as a reasonable lower bound on the density of the inner Uranian satellites. As crossing time increases with decreasing mass, we will use this density to place an upper bound on crossing time.

Running multiple simulations at such a low density would require substantial computation time. Instead, once we have found the power law parameters for the unit-density models, we can extrapolate the results to non-unit-densities because $m_f$ is roughly equivalent to $\rho$ in g/cm$^3$. The resulting predicted crossing times for $\rho = 0.5$ g/cm$^3$ are shown in Table 5. Cupid and Belinda cross orbits on a time scale of $1.0 \times 10^5$ to $1.6 \times 10^7$ years and Cressida and Desdemona cross orbits on a time scale of $1.3 \times 10^6$ to $3.2 \times 10^7$ years, depending on the particular mass assumptions used. Both pairs are likely to collide on time scales significantly shorter than the age of the Uranian system.

*4.4. Interactions with Other Satellites*

Our power law interpretations of the Inner(*sat*±) models also provide insight into the influence of the other satellites on the stability of the Cupid–Belinda and Cressida–Desdemona pairs. By finding significant monotonic changes in crossing time with mass, we can determine which satellites most affect the crossing times of these two pairs. It should be noted, however, that our one-sigma changes in radii are based on observational uncertainties, and thus the percentage mass change for each satellite is different. For example, the three mass models of Cressida only change mass by a total of 34%, while the three mass models of Belinda change mass by nearly a factor of three. Thus it is not appropriate to rank the level of influence of each satellite based on our integrations, only to note that such an influence exists.

We find that the stability of Cupid–Belinda increases with Cupid's mass, but decreases with Belinda's mass. We also find that Perdita has a strong influence, with increasing mass decreasing stability. This is shown in Fig. 6, where the increase in the mass of Perdita, the third smallest satellite in the system, results in a pronounced bend in the power law. Satellites far away from Cupid and Belinda have unexpected influence as well. Bianca appears to have a noticeable effect, with increasing mass resulting in decreased stability. Even more surprising, Mab appears to have an effect, with increasing mass resulting in increased stability. For the Cressida–Desdemona system, the relationships are less clear, perhaps due to the greater difficulty in fitting power law parameters to the secondary pair in our process, or perhaps because the other satellites really do have only a small effect on Cressida and Desdemona. Nevertheless, it appears that Ophelia has an effect on the stability, with increased mass resulting in greater stability.

It is possible that some of these apparent influences (or lack of influences) are artifacts of small-number statistics. To fully quantify the effect of a satellite's mass on system stability, it will be necessary to explore a wider range of masses. Likewise, to properly rank a satellite's influence, it will be necessary to change the satellite's mass in standard increments rather than in increments based on observational uncertainty. These explorations will be the subject of future work.

# 5. LONGER-TERM EVOLUTION

We would like to investigate the further evolution and stability of the Uranian satellites under the conservative mass assumptions established in Section 4.3. What happens after the collisions of Cupid–Belinda and Cressida–Desdemona? To explore this question, we ran simulations with a modified set of satellites. In each case, we used our existing initial conditions except that two or more adjacent satellites were combined, using their initial state vectors, to simulate a perfectly inelastic collision and merger. The semi-major axis of the new, combined satellite was derived from the sum of the orbital energies of the contributing satellites. The remaining orbital elements (eccentricity, inclination, argument of pericenter, longitude of the ascending node, and mean anomaly) were the mass-weighed mean of those from the contributing satellites. The use of the weighted mean for orbital position (e.g. argument of pericenter) is arbitrary. As we are looking at long-term evolution, the exact initial position of a moon in its orbit will not have a significant effect since the power law tends to smooth out the effects of variation due to initial conditions by conducting many simulations. No new orbital fit was performed for the remaining, non-colliding satellites. Although this approach neglects any orbital evolution within the system prior to the time when the pairs of satellites merge, we regard the integrations as generally representative of the system's subsequent evolution.

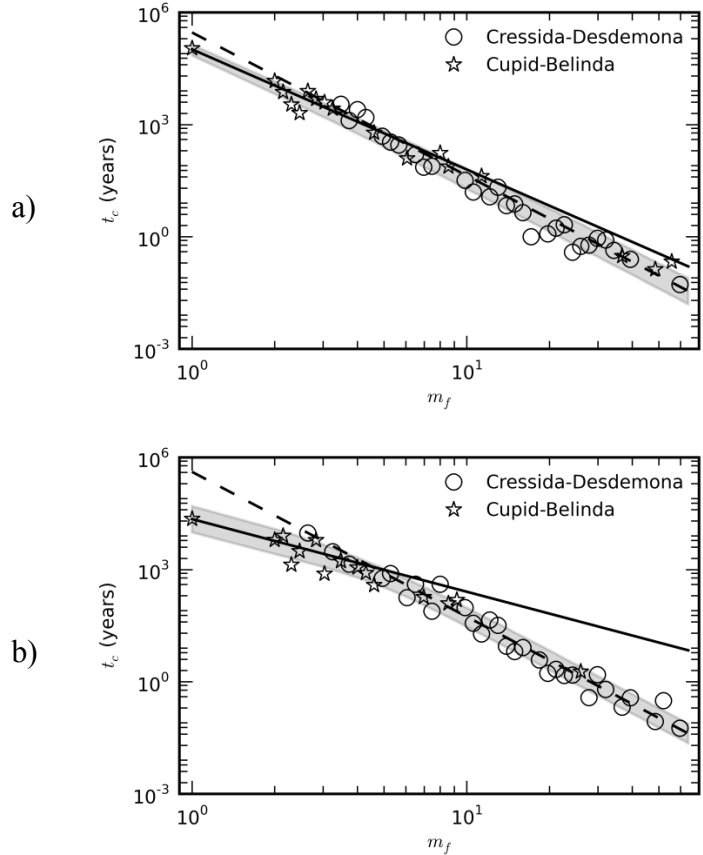

Fig. 6: Dual-power-law fits for the Inner(baseline) model (a) and Inner(Perdita+) model (b). The solid line is the best fit for Cupid–Belinda, the dashed line is the best fit for Cressida–Desdemona, and the ±1σ range of the Monte Carlo results is shown shaded.

To start this process, we examined the 250 position- and mass-perturbed simulations discussed in Section 3.2. For 224 of these simulations, Cupid and Belinda cross orbits first, and thus we feel safe in using this assumption as the root of our cascade. With Cupid and Belinda combined into a single satellite ("CupBel"), and taking advantage of the power law, Cressida and Desdemona are typically the next pair to cross, with log $t_c$ ~ 6.8. We then ran a new series of simulations with both Cupid and Belinda, and Cressida and Desdemona, combined. Because of the increased run times even at higher mass factors, we had to increase the range of mass factors used to 2.828–64. The power law yielded a predicted log crossing time of 9.1. In almost all of these simulations, the combined Cressida–Desdemona satellite (colloquially named "Cresdemona") collided with Juliet.

Continuing the process, CupBel next collided with Perdita (producing "CupBelPer"), with a projected log crossing time of 7.6. Finally, after a combined time of ~1.3 billion years, we are left with a set of satellites where the two most unstable subsets (Cressida–Desdemona–Juliet and Cupid–Belinda–Perdita) have been combined into single satellites. This system took substantially longer to experience crossing events, and we had to once again increase the mass factor range to produce practical simulation times, this time to 32–512. The power law appears to be still valid in this range, yielding a predicted log crossing time of 16.3. While one could argue that such an extrapolation is pushing the power law beyond its probable range of applicability, it is nevertheless evident from such a large result that the system will remain stable for far longer than the remainder of the Uranus system's lifetime.

These estimates closely mimic the predictions presented earlier (Section 3.1) based on Hill sphere separations. If we once again ignore the possible collision of Juliet and Portia, given that empirically they do not cross orbits, and compute the separations with the lower density assumption, the most favorable cascade is: Cupid and Belinda ($\Delta = 12.75$), Cressida and Desdemona ($\Delta = 14.85$), CupBel and Perdita ($\Delta = 23.15$), and Cresdemona and Juliet ($\Delta = 26.53$). This sequence is similar to the empirical sequence above, with the exception that the CupBel and Perdita and the Cresdemona and Juliet collisions are swapped.

## 6. DISCUSSION

All of our numerical integrations show instability and orbit crossings over time scales much shorter than the age of the Uranus system. Cupid is on a particularly precarious orbit, typically crossing the orbit of Belinda within $10^5$ to $10^7$ years. The orbits of Cressida and Desdemona also intersect within $10^6$ to $10^7$ years. These time scales hold for all reasonable assumptions about the masses and densities of the moons in question. Extended integrations suggest that Perdita will subsequently merge with Cupid–Belinda and that Juliet will merge with Cressida–Desdemona in $\sim 10^9$ years. However, the other components of the Uranus system appear to be stable for much longer periods of time. Note that these lifetimes assume a low density of 0.5 g/cm$^3$. If the actual density is higher, the lifetimes will be even shorter.

It is very unlikely for us to be observing the Uranus system so close to the end of the lifetimes of several moons. One possible explanation is that some as yet overlooked phenomenon, such as unidentified resonances, are acting to stabilize the system. However, we regard it as more likely that the inner moons of Uranus exist in a steady state, where new moons are appearing at roughly the same rate that they are being lost.

The existence of the faint ν ring between the orbits of Portia and Rosalind lends support to this hypothesis; Showalter and Lissauer (2006) proposed that this ring represents the debris left behind from an earlier collision. The time scale for such a ring to re-accrete is difficult to determine. Early work on disrupted satellites found that accretion would occur in 10–100 years if tidal forces could be ignored (Canup and Esposito, 1995). If this estimate were correct, then the chance of our seeing an un-accreted ring would be exceedingly small. However, the ν ring is close enough to the Roche zone that tidal forces cannot be neglected. Canup and Esposito (1995) found that, in this region, even perfectly inelastic collisions would only result in accretion under certain circumstances. In particular, the difference in the mass of the two colliding objects needs to be large, because otherwise each object will have significant portions existing outside of the Hill sphere of the other object, and they will not become gravitationally bound. The mass difference requirement causes a bimodal distribution in the size of the resulting bodies as dust fails to aggregate with other dust but successfully combines with larger bodies. Although the dynamical details have not been fully investigated, it would seem plausible that, under these circumstances, the process of re-accretion would be slowed but not halted entirely. If the re-accretion time is more like $\sim 10^5$ to $10^7$ years, then we would expect to see roughly equal numbers of moons on precarious orbits (in this case, two) and the debris clouds from prior collisions (one).

Meteoroid impacts can also disrupt small moons. Colwell et al. (2000) found that catastrophic disruptions of Cressida and Desdemona could occur every 0.9–6.2 Gyr, depending on assumptions about the internal strength of the satellites, while catastrophic disruptions of Belinda could occur every 1.3–9.1 Gyr. Such disruptions have likely already occurred at least once since the formation of the Uranian system. Thus, even if Cressida and Desdemona, or Cupid and Belinda, were able to collide and accrete into a single new body, the resultant body would eventually be broken apart by a subsequent impact; the long-term stability of the inner Uranian satellites discussed in Section 5 will never be achieved. However, the current mean time between catastrophic disruptions appears to be quite a bit longer than the collisional lifetimes we have inferred in this study. Thus, this mechanism does not appear likely to provide a plausible explanation for the recent origin of Cupid.

In the end, there is little concrete we can say about the history or future of the inner Uranian satellites, except that we are not seeing them in their original configuration and what we see today will likewise change over astronomically short periods of time. This evolution will be driven by orbital instabilities and resulting collisions, by the accretion dynamics of colliding bodies and/or their rings of debris, and by the random impacts of large meteoroids.

The moons of Uranus are generally named after characters from the plays of William Shakespeare. Like their namesakes, it appears that several of these moons are also likely to meet tragic and untimely ends.

**Acknowledgments**
We thank the Centre for Astrophysics and Supercomputing at the Swinburne University of Technology along with Professors Jarrod Hurley and Sarah Maddison for access to the Swinburne "Green" supercomputer, which was used for many of the simulations in this paper. We thank Rebekah Dawson, Luke Dones, Alice Quillen, and Matija Ćuk for valuable discussions and comments on this manuscript. This work was supported by NASA's Planetary Geology and Geophysics Program under grant NNX09AG14G. Additional support for Program AR-10977.01 was provided by NASA through a grant from the Space Telescope Science Institute, which is operated by the Association of Universities for Research in Astronomy, Incorporated, under NASA contract NAS5-26555.